\documentclass[letterpaper]{jpconf}
\usepackage{graphicx}
\begin{document}
\title{Heavy Flavour Physics at CDF}

\author{Gavril Giurgiu (for the CDF collaboration)}

\address{Johns Hopkins University}

\begin{abstract}
We present recent CDF measurements of mass, lifetime and CP 
violation properties of B hadrons.
The analyzes presented in this paper use up to 2.8~fb$^{-1}$ of data. 
CDF has already accumulated close to 5~fb$^{-1}$ of data which promises significant 
improvements of these analyzes in the near future.  
\end{abstract}

\section{Introduction}

We present recent Heavy Flavour Physics results from the CDF experiment~\cite{cdf_detector}  
which has accumulated 5~fb$^{-1}$ of data. The measurements presented in this paper 
make use of up to 2.8~fb$^{-1}$ of data. We summarize results of the $B_c$ mass and lifetime, 
$\Lambda_b$ and $B_s$ lifetimes measured for the first time from samples selected 
by a trigger which requires the presence of tracks with large impact parameter with respect 
to the primary vertex and CP violation in $B_s \rightarrow J/\psi \phi$ decays.

\section{Mass and Lifetime of the $B_c$ Meson }

The $B_c$ meson is unique as it contains two heavy quarks: $b$ (bottom) and 
$\bar{c}$ (anti-charm). 
The most precise measurement of the $B_c$ mass was recently performed by the CDF 
experiment with $2.4~$fb$^{-1}$ of data \cite{Bc_mass_CDF_run_II} using the fully reconstructed 
decay mode $B_c^+ \rightarrow J/\psi [\rightarrow \mu^+ \mu^-] \pi^+$. The advantage of using fully reconstructed decays 
is that the mass of the decaying particle can be measured precisely by fitting 
the invariant mass distribution. 
The total $B_c$ signal yield is estimated to $108 \pm 15$ candidates 
with a statistical significance of 8 standard deviations. The measured mass is
$6275.6 \pm 2.9(stat.) \pm 2.5(syst.)$~MeV/c$^{2}$ which is good agreement with 
the corresponding D0 measurement~\cite{Bc_mass_D0} and with theoretical 
models which predict the $B_c$ mass around $6.3$~GeV/c$^2$~\cite{Bc_mass_potential}. 

The lifetime of the $B_c$ meson is measured by the CDF experiment using partially 
reconstructed $B_c \rightarrow J/\psi [\rightarrow \mu^+ \mu^-]\,l\,\nu\,X$ decays \cite{Bc_lifetime_CDF} 
in $1.0$~fb$^{-1}$ of data. Measurements are performed separately in $J/\psi\,e$ and $J/\psi\,\mu$ 
channels. 
The corresponding lifetimes are: $c\tau_{\mu} = 179.1 ^{+32.6}_{-27.2}(stat.)$ 
and $c\tau_{e} = 121.7 ^{+18.0}_{-16.3}(stat.)\,\mu$m.     
The main sources of systematic uncertainties come from our understanding of the 
decay time resolution function and from relative fractions of $b \bar{b}$ 
production mechanisms in simulation. The total systematic uncertainty is $5.5\,\mu$m, 
leading to a combined measurement of the $B_c$ lifetime of 
$c\tau_{\mu} = 142.5 ^{+15.8}_{-14.8}(stat.) \pm 5.5(syst.)\,\mu$m. 
This agrees well with a similar recent measurement performed by the D0 experiment \cite{Bc_lifetime_D0} 
in the $B_c \rightarrow J/\psi \mu~\nu~X$  channel 
and with theoretical predictions \cite{Bc_lifetime}. 

\section{Lifetime of the $\Lambda_b$ Baryon}

\begin{figure}[htb]
\vspace{9pt}
\includegraphics[width=50mm]{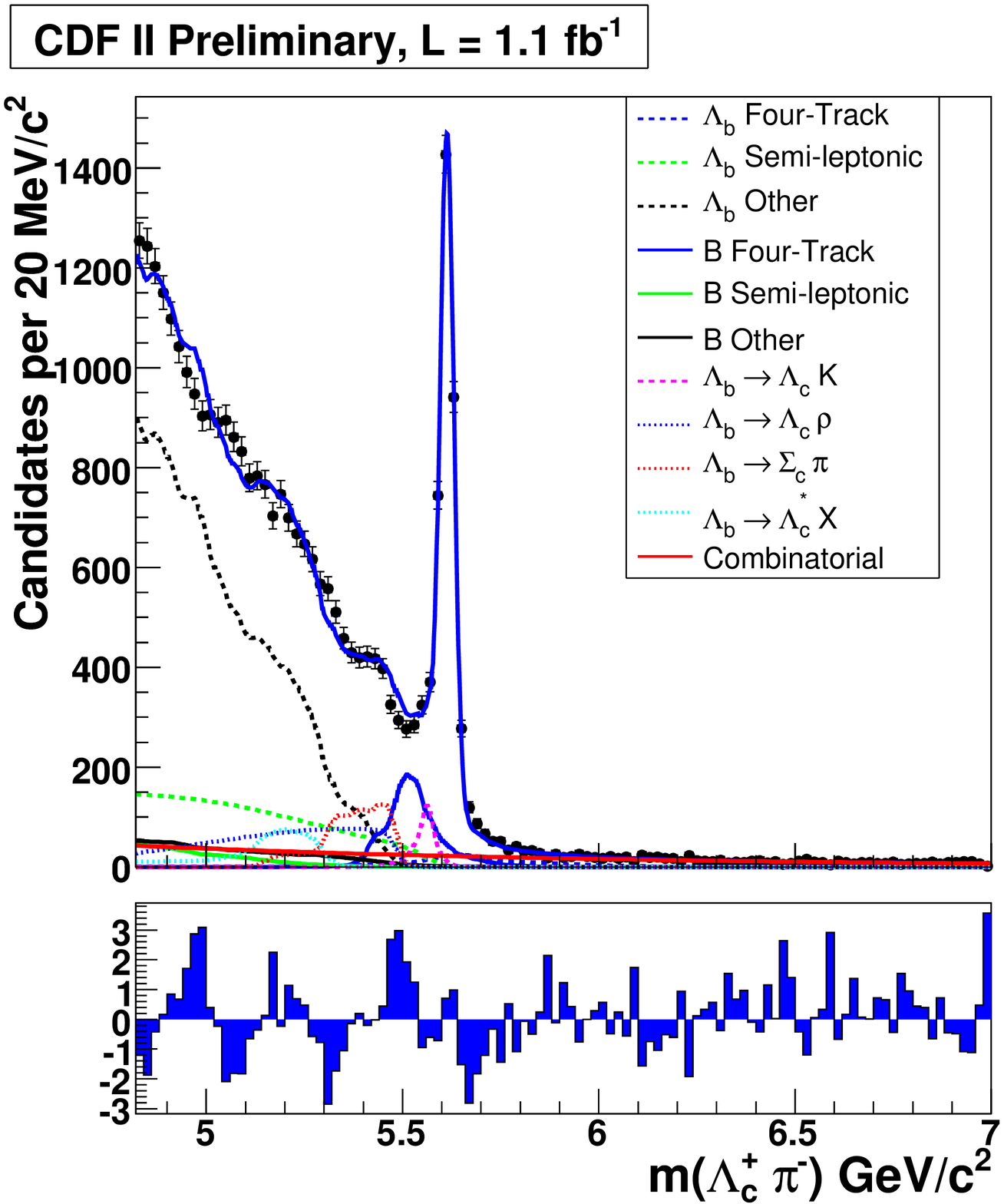}
\includegraphics[width=50mm]{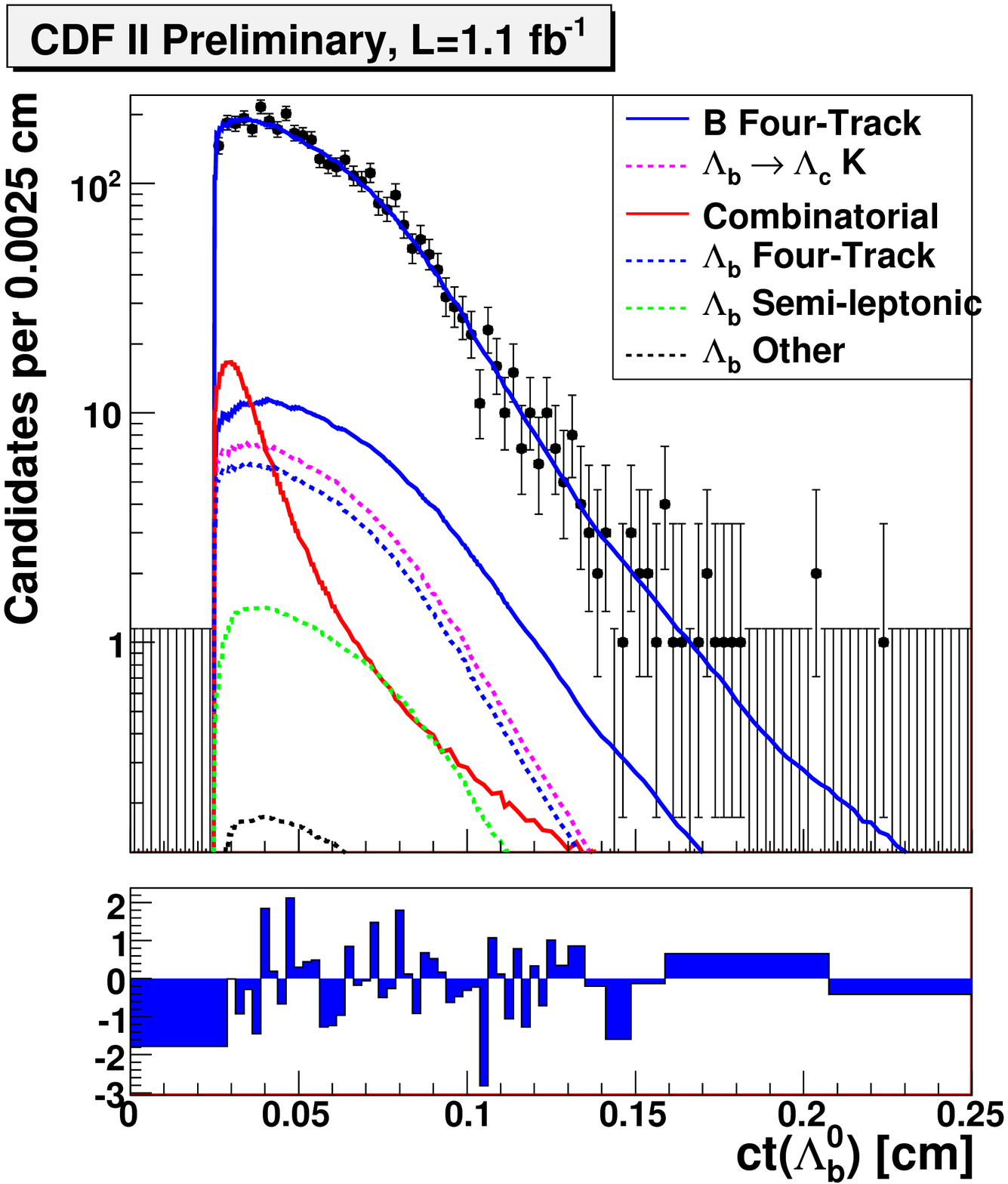}
\includegraphics[width=60mm]{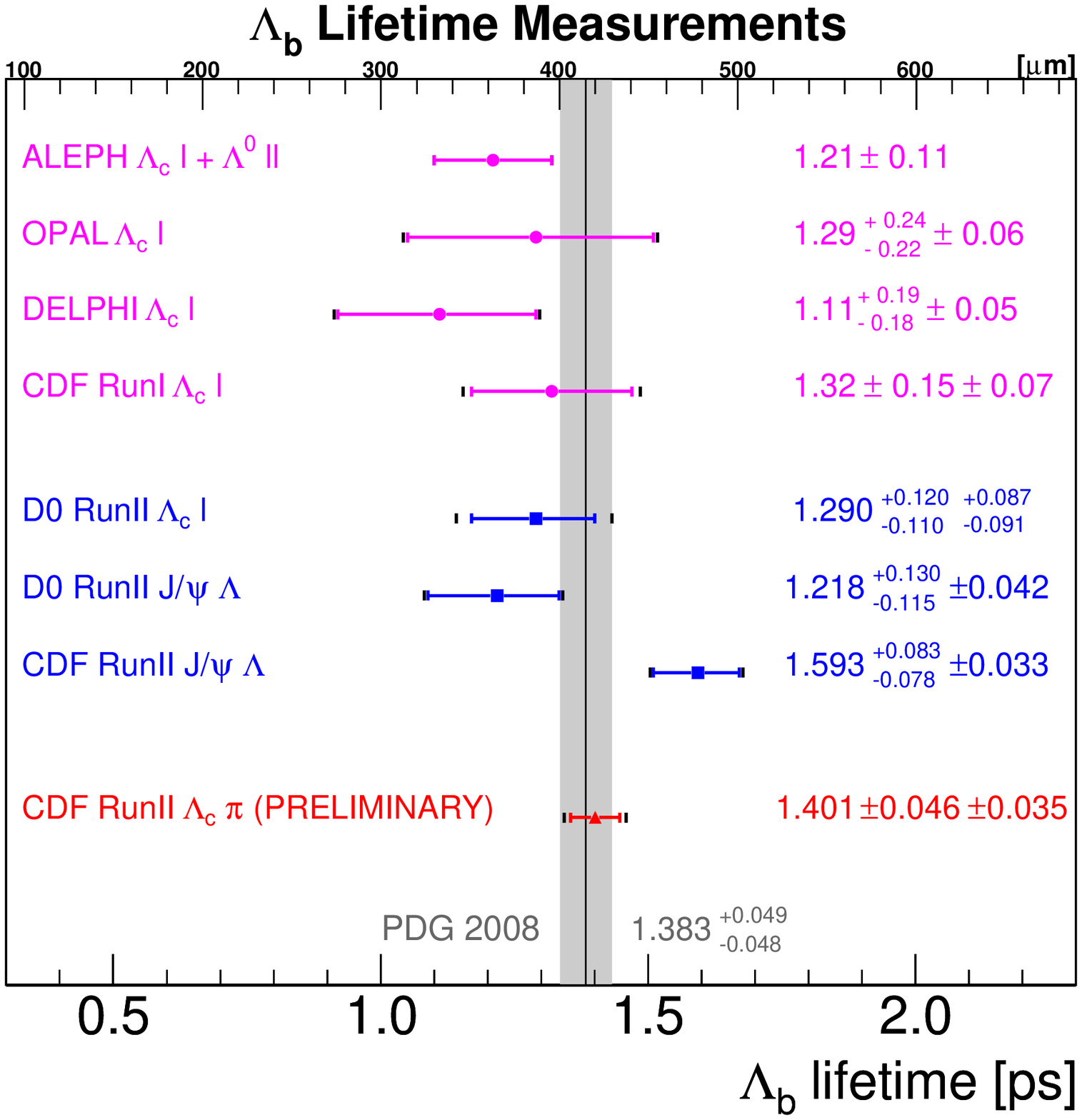}
\caption{ Invariant $\Lambda_c \pi$ mass (left) and $\Lambda_b$ decay time (center) 
distributions with fit projections superimposed. Comparison between 
different $\Lambda_b$ lifetime measurements and the current world average (right). }
\label{fig:lambda_b_mass}
\end{figure}

The lifetime of the $\Lambda_b$ baryon was measured for the first time in a fully hadronic decay 
mode, $\Lambda_b^0 \rightarrow \Lambda_c^+ [\rightarrow p^+ K^- \pi^+] \pi^-$, by the CDF experiment. 
This was made possible by using a  
trigger which selects events with two tracks with large impact parameter ($d_0$) with respect to 
the primary vertex, $d_0 > 120 \mu$m. The large impact parameter of secondary vertex  
tracks is characteristic to long lived $b$ hadrons like $\Lambda_b$ and it is a powerful discriminant 
against prompt background. This analysis uses 1.1 fb$^{-1}$ 
of data in which a total of 3000 signal $\Lambda_b$ events are found after signal to background 
optimization. The invariant $\Lambda_b$ mass and  
decay time distributions are shown in Fig~\ref{fig:lambda_b_mass}.
Because the sample is selected by a trigger which requires displaced tracks, events 
with low decay time are suppressed. The trigger efficiency is determined from simulation 
and used in the maximum likelihood fit to correct for this effect. The measured 
lifetime is~\cite{Lb_lifetime} 
$c\tau(\Lambda_b) = 420.1 \pm 13.7(stat.) \pm 10.6 (syst.) \mu$m. The main systematic 
uncertainty comes from our limited knowledge the trigger efficiency simulation. The 
ratio between the $\Lambda_b$ lifetime measured in this analysis and the $B^0$ world 
average lifetime~\cite{pdg} is $0.922 \pm 0.039$ which is in good agreement with 
theoretical calculations~\cite{Lb_lifetime_th} which predict a ratio of $0.88 \pm 0.05$ 
and with previous results summarized in Fig.~\ref{fig:lambda_b_mass}.  
         
\section{Lifetime of the $B_s$ Meson}

\begin{figure}[htb]
\vspace{9pt}
\includegraphics[width=50mm]{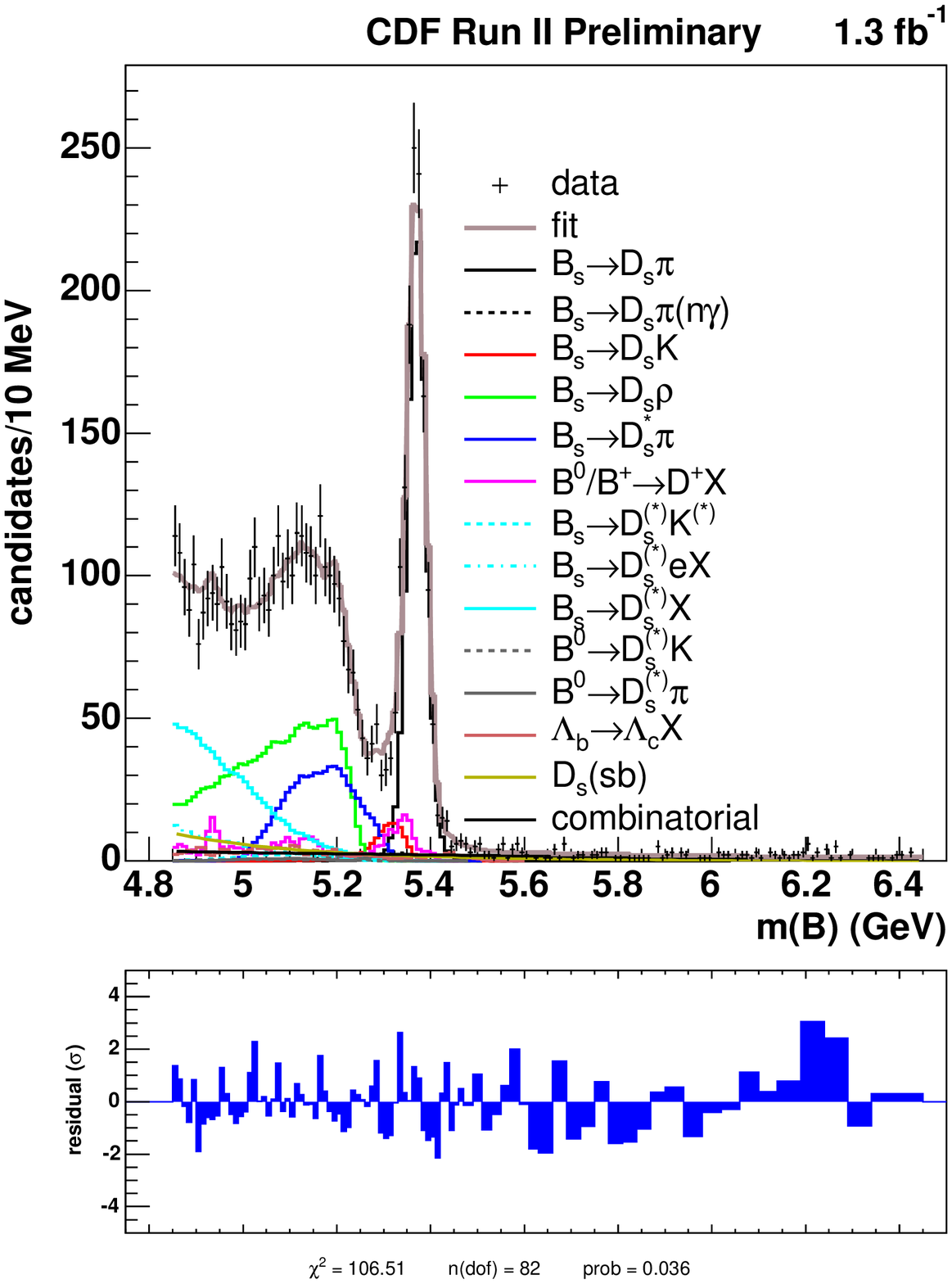}
\includegraphics[width=50mm]{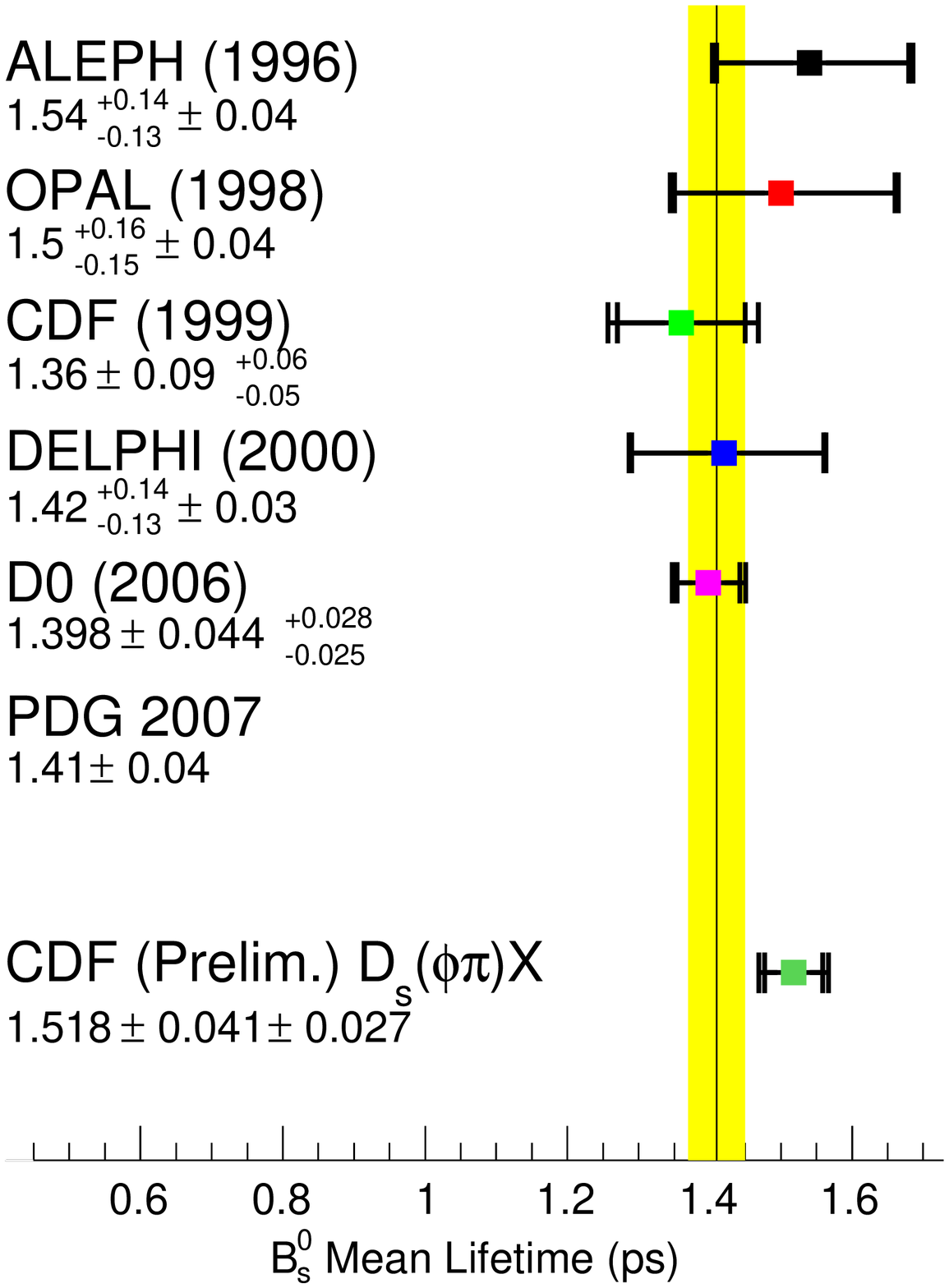}
\includegraphics[width=70mm]{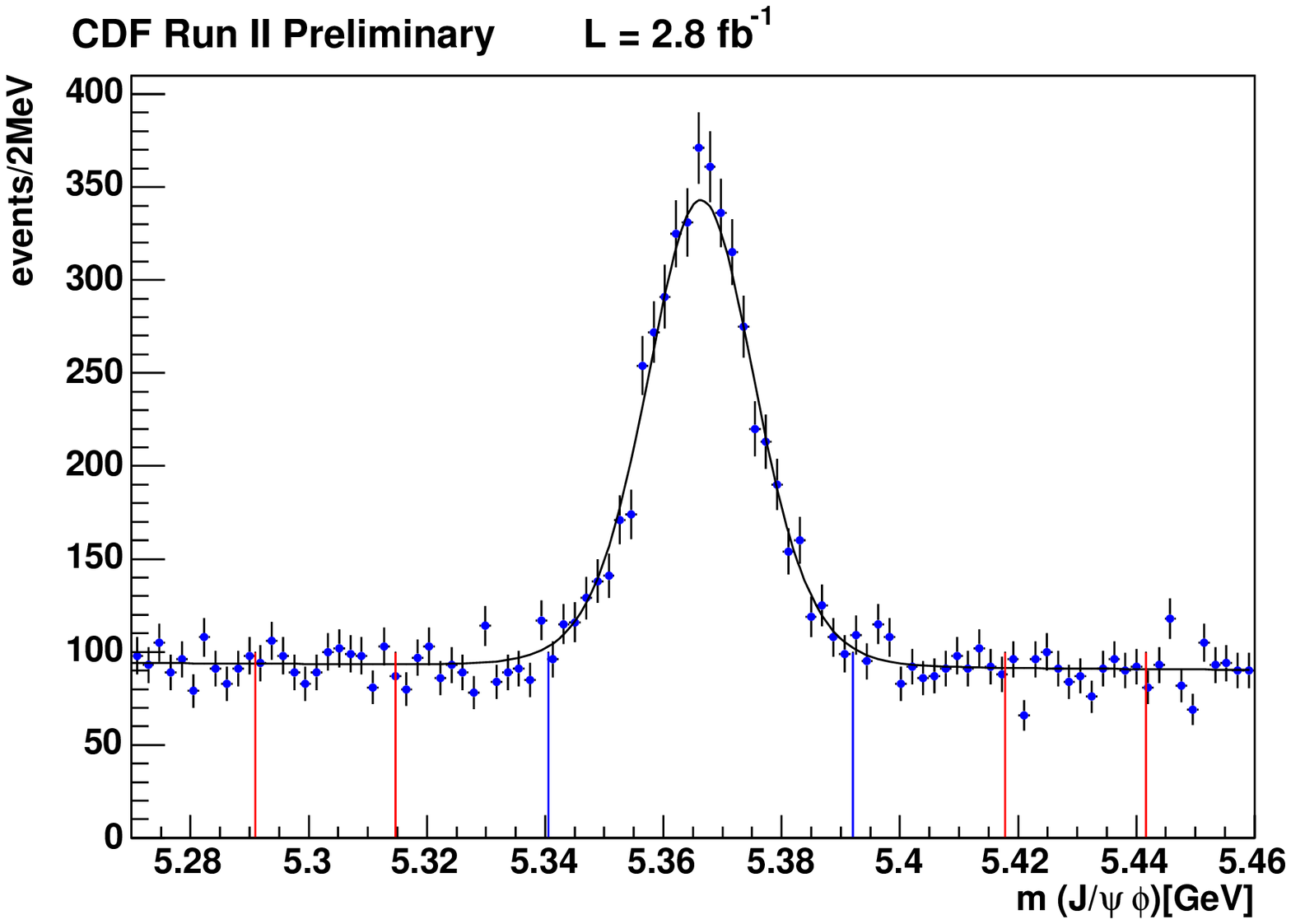}
\caption{ Invariant $D_s \pi$ mass distribution with fit projections superimposed (left). Comparison between the 
recent CDF $B_s$ lifetime measurement and previous measurements (center). Invariant $J/\psi \phi$ mass
distribution with fit projection superimposed (right). }
\label{fig:bs_mass}
\end{figure}

The CDF experiment has performed for the first time a measurement of the $B_s$ lifetime in  
fully hadronic $B_s^0 \rightarrow D_s^- [\rightarrow \phi \pi^-] \pi^+$ decays, where $\phi \rightarrow K^+ K^-$. 
Since the $B_s$ flavor ($B_s$ or $\bar{B}_s$) can be inferred from the decay products, this decay 
mode is refereed to as flavor specific.    
The analysis also includes partially reconstructed decays like  
$B_s \rightarrow D_s^- \rho^+ [\rightarrow \pi^+ \pi^0]$ where the $\pi^0$ from 
the $\rho$ decay is not reconstructed. This analysis uses 1.3 fb$^{-1}$ of data which  
was collected by the same displaced track trigger used for the $\Lambda_b$ 
lifetime measurement described in the previous section. A similar trigger 
bias correction is applied. 
After signal to background optimization the sample yields 1100 fully reconstructed 
$B_s$ decays and about twice as many partially reconstructed $B_s$ decays. 
The $D_s \pi$ invariant mass distribution is shown in Fig~\ref{fig:bs_mass}.
The measurement procedure is tested on larger control samples of $B^0 \rightarrow D^- \pi^+$, 
$B^0 \rightarrow D^{*-} \pi^+$ and $B^+ \rightarrow \bar{D}^0 \pi^+$ decays, where the 
$B^+$ and $B^0$ lifetimes are measured in good agreement with the world averages~\cite{pdg}. 
This analysis~\cite{Bs_lifetime} yields the best $B_s$ lifetime in flavor specific decays 
$\tau(B_s) = 1.518 \pm 0.041 (stat.) \pm 0.025 (syst.)$. As in the case of the 
$\Lambda_b$ lifetime measurement, the largest systematic uncertainty comes 
from the limited understanding of the trigger simulation. The $B_s$ lifetime 
measured in this analysis is larger than the previous world average as seen in 
Fig~\ref{fig:bs_mass}. This new result will push the previous world average ratio between 
the $B_s$ lifetime and the $B^0$ lifetime ($0.94 \pm 0.02$) to a larger value, closer to the 
Heavy Quark Effective Theory~\cite{hqet} prediction of $1.0 \pm 0.02$.

\section{CP Violation in $B_s \rightarrow J/\psi \phi$ Decays}

A meson with $B_s^0$ flavor at production time can have the same flavor $B_s^0$ at decay time or 
it can decay from an oscillated $\bar{B}_s^0$ state. This is possible because $B_s^0$ and $\bar{B}_s^0$ mesons 
oscillate from one another with high frequency. The relative phase between the two possible decay 
amplitudes, $\beta_s = \arg\left(-\frac{V_{ts}V_{TB}^*}
{V_{cs}V_{cb}^*}\right)$ is responsible 
for CP violation in $B_s \rightarrow J/\psi \phi$ decays. In the Standard Model (SM) this phase is predicted to be 
very small $O(\lambda^2) \approx 0.02$~\cite{Ref:lenz}.  
Although the final state in this decay, $J/\psi [\rightarrow \mu^+ \mu^-] \phi [\rightarrow K^+ K^-]$, is not a 
CP eigenstate, it is however a superposition of CP eigenstates. Since both $J/\psi$ and $\phi$ are spin 1 
particles, the final state $J/\psi \phi$ has total angular momentum of either 0, 1 or 2. The states 
with angular momentum 0 and 2 are CP even while the state with angular momentum 1 is CP odd. 
The CP even and CP odd eigenstates can be statistically separated by doing an angular analysis 
of the final decay products $\mu^+ \mu^-$ and $K^+ K^-$. Both same side and opposite side flavor 
tagging techniques are used to separate $B_s^0$ from $\bar{B}_s^0$ production flavors.       

\begin{figure}[htb]
\vspace{9pt}
\includegraphics[width=65mm]{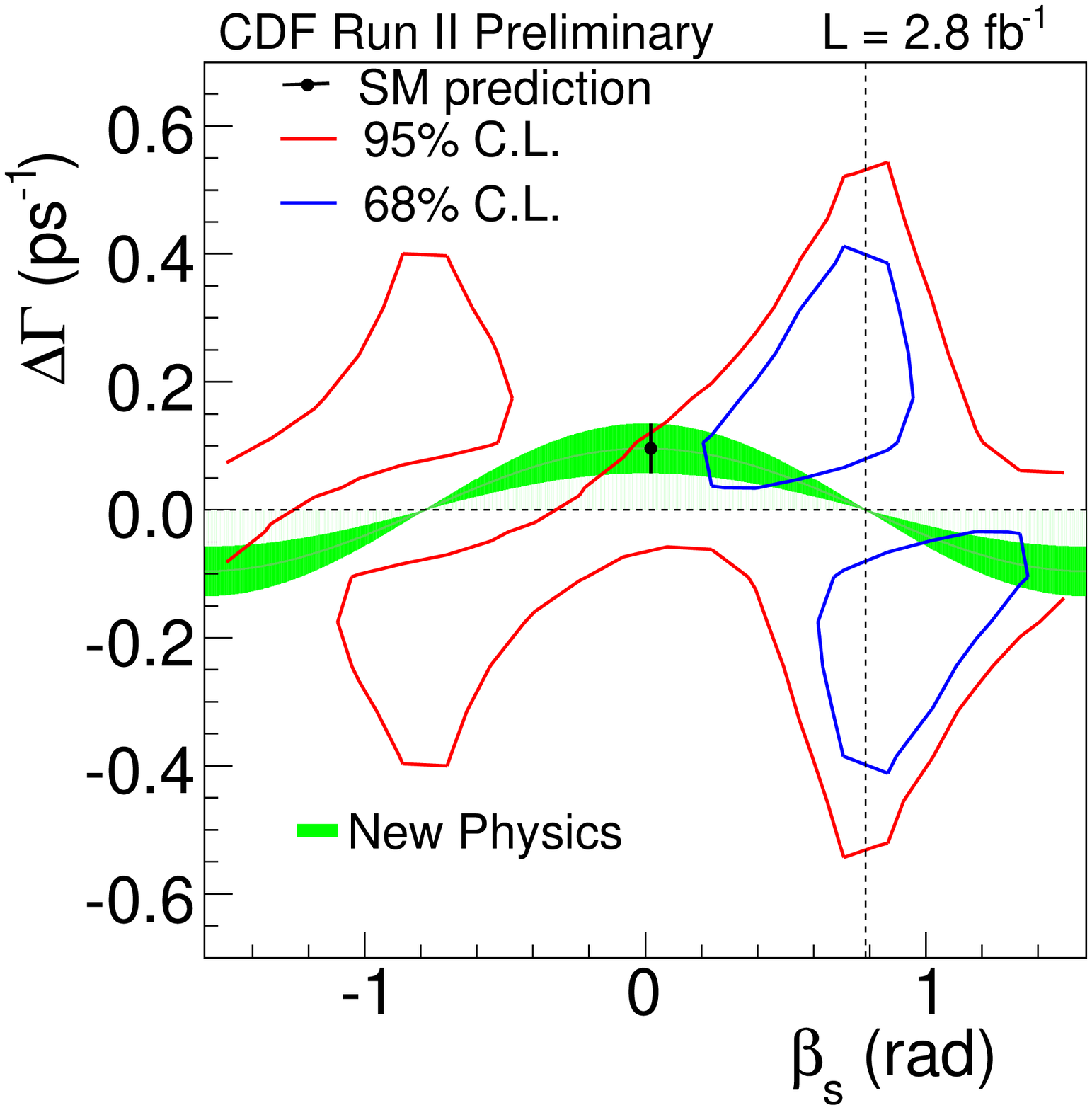}
\includegraphics[width=75mm]{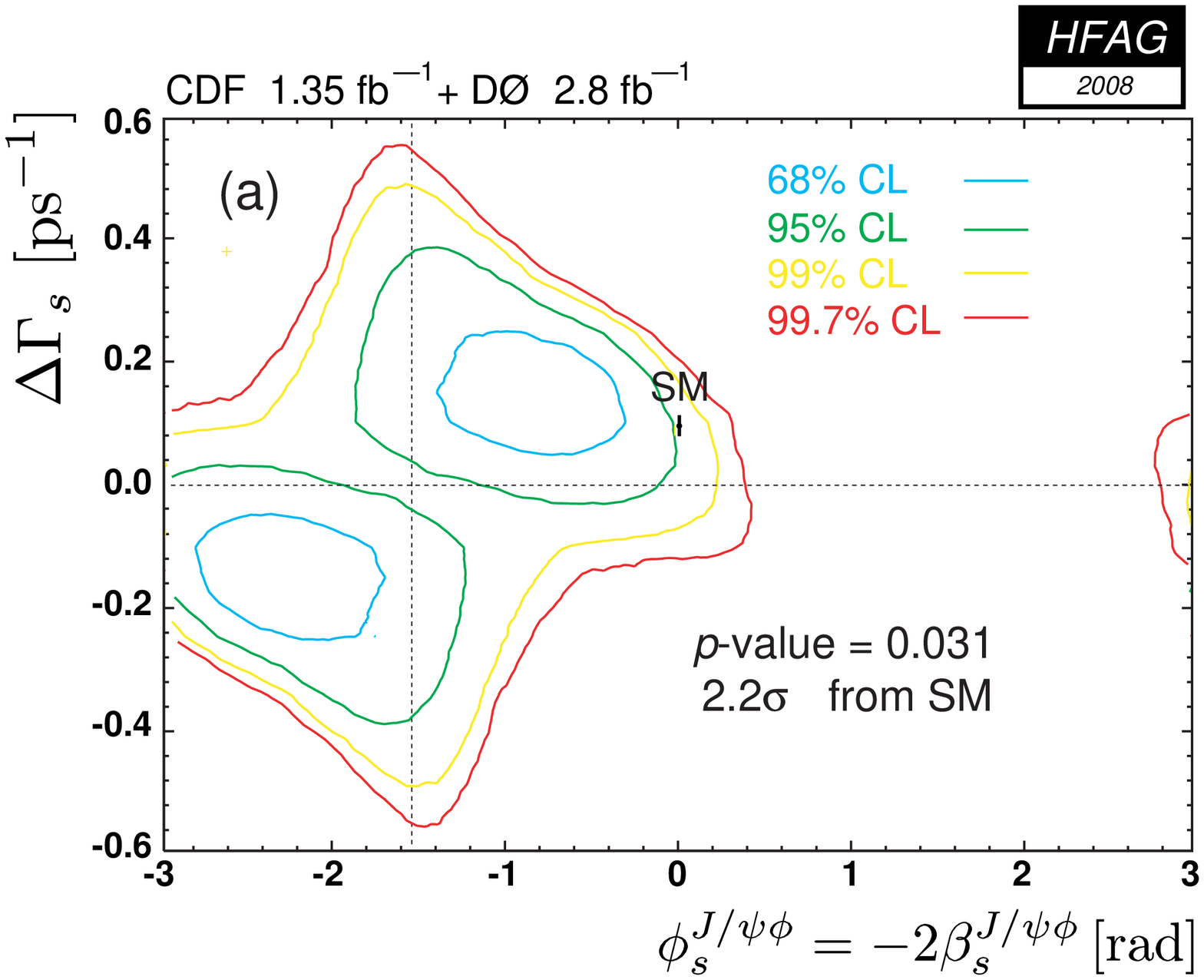}
\caption{ Confidence regions in the $\beta_s^{J/\psi\,\phi}-\Delta\Gamma_s$ plane from CDF (left) and 
	in the $\phi_s^{J/\psi\,\phi}-\Delta\Gamma_s$ plane from combined CDF and D0 datasets (right). 
	Note the definition $\phi_s^{J/\psi\,\phi} = -2 \beta_s^{J/\psi\,\phi}$. }
\label{fig:2d_contours}
\end{figure}

The CDF experiment has performed a study~\cite{cdf_beta_s} of CP violation in 
$B_s \rightarrow J/\psi \phi$ decays using 2.8~fb$^{-1}$ of data selected by a di-muon trigger. After signal optimization  
the sample contains about 3200 $B_s \rightarrow J/\psi \phi$ signal events as seen in Fig~\ref{fig:bs_mass}. 
The measured average $B_s$ lifetime is $1.53 \pm 0.04 (stat.) \pm 0.01 (syst.)$~ps. 
Assuming no CP violation the decay width difference between the heavy and light $B_s$ 
mass eigenstates is determined $\Delta \Gamma_s = 0.02 \pm 0.05 (stat.) \pm 0.01 (syst.)$~ps$^{-1}$. 
When allowing for CP violation in the maximum likelihood fit, non-Gaussian effects 
are accounted for by using a frequentist approach for determining two dimensional 
confidence regions in the $\beta_s - \Delta \Gamma_s$ plane. These confidence regions 
are shown in Fig.~\ref{fig:2d_contours}.
This analysis provides strong indication 
that negative values of the CP violation phase $\beta_s$ are suppressed. The agreement 
with the $\beta_s$ and $\Delta \Gamma_s$ SM prediction~\cite{Ref:lenz} is  
$7\%$ or 1.8 standard deviations. A similar masurement carried out 
by the D0 experiment~\cite{d0_beta_s} leads to similar conclusions: negative values of 
$\beta_s^{J/\psi\,\phi}$ are disfavoured while the agreement with the 
SM expectation is 1.7 standard deviations. A combination of an older  
CDF measurement~\cite{cdf_beta_s_initial} and the D0 measurement~\cite{d0_beta_s} has been 
performed by the Heavy Flavor Averaging Group~\cite{Ref:HFAG}. 
The combined confidence regions are shown in Fig.~\ref{fig:2d_contours}. The 
agreement with the SM prediction is at the $3\%$ level or 2.2 standard 
deviations. It will be interesting to see the evolution of these 
results as the Tevatron experiments are expected to accumulate between 6 and 8~fb$^{-1}$ 
by the end of Run 2. 

%
  
\section{Conclusions}
   
The CDF experiment has a very rich B Physics program which is complementary to, 
and in some aspects competitive with the B factories. Some of the best 
measurements of the $B_c$, $\Lambda_b$ and $B_s$ hadron properties. 
Updated analyzes will 
follow as CDF is expected to accumulate 6 to 8~fb$^{-1}$ of data by the 
end of the Tevatron running.        
 

\end{document}